\documentclass[a4paper,UKenglish]{lipics}
 
\usepackage{microtype}

\title{NaDeA: A Natural Deduction Assistant with a Formalization in Isabelle}
\author{J{\o}rgen Villadsen}
\author{Alexander Birch Jensen}
\author{Anders~Schlichtkrull}
\affil{DTU Compute - Department of Applied Mathematics and Computer Science, Technical University of Denmark, Richard Petersens Plads, Building 324, DK-2800 Kongens Lyngby, Denmark \\ \texttt{jovi@dtu.dk}}

\authorrunning{J. Villadsen, A.\,B. Jensen and A. Schlichtkrull}

\Copyright{J{\o}rgen Villadsen, Alexander Birch Jensen \&\ Anders Schlichtkrull}


\serieslogo{logo_ttl}
\volumeinfo
  {M. Antonia {Huertas}, Jo\~ao {Marcos}, Mar\'ia {Manzano}, Sophie {Pinchinat}, \\
  Fran\c{c}ois {Schwarzentruber}}
  {5}
  {4th International Conference on Tools for Teaching Logic}
  {1}
  {1}
  {253}
\EventShortName{TTL2015}

\newcommand{\x}[1]{\textsf{#1}}

\newcommand{\nadea}{\x{NaDeA}}

\newcommand{\conwidthx}{3.7ex}
\newcommand{\uniwidthx}{3.3ex}
\newcommand{\qmwidthx}{1ex}

\usepackage{proof}

\begin{document}

\maketitle

\begin{abstract}
We present a new software tool for teaching logic based on natural deduction. Its proof system is formalized in the proof assistant Isabelle such that its definition is very precise. Soundness of the formalization has been proved in Isabelle. The tool is open source software developed in TypeScript / JavaScript and can thus be used directly in a browser without any further installation. Although developed for undergraduate computer science students who are used to study and program concrete computer code in a programming language we consider the approach relevant for a broader audience and for other proof systems as well.
\end{abstract}

\section{Introduction}

In this paper we present the \nadea\ software tool. First, we provide the motivation and a short description. We then present the natural deduction system as it is done in a popular textbook \cite{Huth} and as is it done in \nadea\ by looking at its formalization in Isabelle. This illustrates the differences between the two approaches. We also present the semantics of first-order logic as formalized in Isabelle, which was used to prove the proof system of \nadea\ sound. Thereafter we explain how \nadea\ is used to construct a natural deduction proof. Lastly, we compare \nadea\ to other natural deduction assistants and consider how \nadea\ could be improved.

\subsection{Motivation}

We have been teaching a bachelor logic course --- with logic programming --- for a decade using a textbook with emphasis on tableaux and resolution \cite{Ben-Ari}.
We have started to use the proof assistant Isabelle \cite{Nipkow} and refutation proofs are less preferable here.
The proof system of natural deduction \cite{Prawitz,Pelletier,Fitting,Huth} with the introduction and elimination rules as well as a discharge mechanism seems more suitable.
The natural deduction proof system is widely known, used and studied among logicians throughout the world.
However, our experience shows that many of our undergraduate computer science students struggle to understand the most difficult aspects.

This also goes for other proof systems. The formal language of logic can be hard to teach our students because they do not have a strong theoretical mathematical background. Instead, most of the students have a good understanding of concrete computer code in a programming language.
The syntax used in Isabelle is in many ways similar to a programming language, and therefore a clear and explicit formalization of first-order logic and a proof system may help the students in understanding important details. Formalizations of model theory and proof theory of first-order logic are rare, for example \cite{Harrison,Berghofer,Blanchette}.

\subsection{The Tool}

We present the natural deduction assistant \nadea\ with a formalization in the proof assistant Isabelle of its proof system.
It can be used directly in a browser without any further installation and is available here: 
\begin{center}
\url{http://nadea.compute.dtu.dk/}
\end{center}
\nadea\ is open source software developed in TypeScript / JavaScript and stored on GitHub. The formalization of its proof system in Isabelle is available here:
\begin{center}
\url{http://logic-tools.github.io/}
\end{center}
Once \nadea\ is loaded in the browser --- about 250 KB with the jQuery Core library --- no internet connection is required. Therefore \nadea\ can also be stored locally.

We display the natural deduction proofs in two different formats. We present the proof in an explicit code format that is equivalent to the Isabelle syntax, but with a few syntactic differences to make it easier to understand for someone trying to learn Isabelle. In this format, we present the proof in a style very similar to that of Fitch's diagram proofs. We avoid the seemingly popular Gentzen's tree style to focus less on a visually pleasing graphical representation that is presumably much more challenging to implement.

We find that the following requirements constitute the key ideals for any natural deduction assistant. It should be:

\begin{itemize}
\item[--] Easy to use.
\item[--] Clear and explicit in every detail of the proof.
\item[--] Based on a formalization that can be proved at least sound, but preferably also complete.
\end{itemize}

Based on this, we saw an opportunity to develop \nadea\ which offers help for new users, but also serves to present an approach that is relevant to the advanced users.

In a paper considering the tools developed for teaching logic over the last decade \cite[p.~137]{Huertas}, the following is said about assistants
(not proof assistants like Isabelle but tools for learning/teaching logic):
\begin{quotation}
\noindent
Assistants are characterized by a higher degree of
interactivity with the user. They provide menus and dialogues to the user for interaction purposes. This kind of tool gives the students the feeling that they are being helped in building the solution. They provide error messages and hints in the guidance to the construction of the answer. Many of them usually offer construction of solution in natural deduction proofs. [...] They are usually free licensed and of open access.
\end{quotation}
We think that this characterization in many ways fits \nadea. While \nadea\ might not bring something new to the table in the form of delicate graphical features, we emphasize the fact that it has some rather unique features such as a formalization of its proof system in Isabelle. 

\section{Natural Deduction in a Textbook}

We consider natural deduction as presented in a popular textbook on logic in computer science \cite{Huth}. First, we take a look substitution, which is central to the treatment of quantifiers in natural deduction.

\subsection{On Substitution}

The following definition for substitution is used in \cite[p.~105~top]{Huth}:
\begin{quotation}
\noindent
Given a variable $x$, a term $t$ and a formula $\phi$ we define $\phi[t/x]$ to be the formula obtained by replacing each free occurrence of variable $x$ in $\phi$ with $t$.
\end{quotation}

The usual side conditions that come with rules using this substitution seem to be omitted.
In \cite[p.~106~top]{Huth}, we are shortly after given the following definition of what it means that '$t$ must be free for $x$ in $\phi$':
\begin{quotation}
\noindent
Given a term $t$, a variable $x$ and a formula $\phi$, we say that $t$ is free for $x$ in $\phi$ if no free $x$ leaf in $\phi$ occurs in the scope of $\forall y$ or $\exists y$ for any variable $y$ occurring in $t$.
\end{quotation}

The following quote \cite[p.~106~bottom]{Huth} from the same book emphasizes how it seems more preferable, due to the high level of complexity, to avoid the details of these important side conditions:
\begin{quotation}
\noindent
It might be helpful to compare '$t$ is free for $x$ in $\phi$' with a precondition of calling a procedure for substitution. If you are asked to compute $\phi[t/x]$ in your exercises or exams, then that is what you should do; but any reasonable implementation of substitution used in a theorem prover would have to check whether $t$ is free for $x$ in $\phi$ and, if not, rename some variables with fresh ones to avoid the undesirable capture of variables.
\end{quotation}

We find that this way of presenting natural deduction proof systems leaves out some important notions that the students ought to learn. In our formalization such notions and their complications become easier to explain because all side conditions of the rules are very explicitly stated. We see it as one of the major advantages of presenting this formalization to students.

\subsection{Natural Deduction Rules}

We now present the natural deduction rules as described in the literature, again using \cite{Huth}. The first 9 are rules for classical propositional logic and the last 4 are for first-order logic.
Intuitionistic logic can be obtained by omitting the rule \textit{PBC} (proof by contradiction, called ``Boole'' later) and adding the $\bot$-elimination rule (also known as the rule of explosion) \cite{Seldin}. 
The rules are as follows:
\begin{trivlist}
\item
$$
\infer[\textit{PBC}]
{	
	~~
    \phi
	~~
}
{
	~~
	\boxed{ 
		\deduce[\raisebox{1.5ex}{\vdots}]{\bot}{\neg \phi}
	}
	~~   
}
~~~~~~~~~~~~~~~~
\infer[\rightarrow E]
{
	~~
	\psi
	~~
}    
{
	~~
	\phi
    ~~&~~
    \phi \rightarrow \psi
	~~   
}
~~~~~~~~~~~~~~~~
\infer[\rightarrow I]
{
	~~
	\phi \rightarrow \psi
 	~~
}
{
	~~
	\boxed{ 
		\deduce[\raisebox{1.5ex}{\vdots}]{\psi}{\phi}
	}
	~~   
}
$$
\par
$$
\infer[\vee E]
{
	~~
	\chi
 	~~
}
{
	~~
    \phi \vee \psi
    ~~&~~
	\boxed{ 
		\deduce[\raisebox{1.5ex}{\vdots}]{\chi}{\phi}
	}
    ~~&~~    
	\boxed{ 
		\deduce[\raisebox{1.5ex}{\vdots}]{\chi}{\psi}
	}
	~~   
}
~~~~~~~~~~~~~~~~
\infer[\vee I_1]
{
	~~
	\phi \vee \psi
 	~~
}
{
	~~
	\phi
	~~   
}
~~~~~~~~~~~~~~~~
\infer[\vee I_2]
{
	~~
	\phi \vee \psi
 	~~
}
{
	~~
	\psi
	~~   
}
$$
\par
$$
\infer[\land E_1]
{
	~~
	\phi
 	~~
}
{
	~~
	\phi \wedge \psi
	~~   
}
~~~~~~~~~~~~~~~~
\infer[\land E_2]
{
	~~
	\psi
 	~~
}
{
	~~
	\phi \wedge \psi
	~~   
}
~~~~~~~~~~~~~~~~
\infer[\land I]
{
	~~
	\phi \wedge \psi
 	~~
}
{
	~~
	\phi
    ~~&~~
    \psi
	~~   
}
$$
\par
$$
\infer[\exists E]
{
	~~
	\chi
 	~~
}
{
	~~
	\exists x\,\phi
    ~~&~~
	\boxed{ 
		\deduce[\raisebox{1.5ex}{~~~~\vdots}]{~~~~\chi}{x_0~~~~\phi\left[x_0/x\right]}
	}
	~~~~~~~~
}
~~~~~~~~~~~~~~~~
\infer[\exists I]
{
	~~
	\exists x\,\phi
 	~~
}
{
	~~
	\phi\left[t/x\right]
	~~   
}
$$
\par
$$
\infer[\forall E]
{
	~~
	\phi\left[t/x\right]
 	~~
}
{
	~~
	\forall x\,\phi
	~~   
}
~~~~~~~~~~~~~~~~
\infer[\forall I]
{
	~~
	\forall x\,\phi
 	~~
}
{
	~~
	\boxed{ 
		\deduce[\raisebox{1.5ex}{~~~~\vdots}]{~~~\phi\left[x_0/x\right]}{x_0~~~~~~~~~~~~}
	}
	~~   
}
$$
\end{trivlist}
Side conditions to rules for quantifiers:
\begin{trivlist}
\item $\exists E$: $x_0$ cannot occur outside its box (and therefore not in $\chi$).
\item $\exists I$: $t$ must be free for $x$ in $\phi$.
\item $\forall E$: $t$ must be free for $x$ in $\phi$.
\item $\forall I$: $x_0$ is a new variable which does not occur outside its box.
\end{trivlist}
In addition there is a special copy rule \cite[p.~20]{Huth}:

\begin{quotation}
\noindent
A final rule is required in order to allow us to conclude a box with a formula which has already appeared earlier in the proof. [...] The rule `copy' allows us to repeat something that we know already. We need to do this in this example, because the rule $\rightarrow I$ requires that we end the inner box with $p$. The copy rule entitles us to copy formulas that appeared before, unless they depend on temporary assumptions whose box has already been closed. Though a little inelegant, this additional rule is a small price to pay for the freedom of being able to use premises, or any other `visible' formulas, more than once.
\end{quotation}
The copy rule is not needed in our formalization due to the way it manages assumptions.

As it can be seen, there are no rules for truth, negation or biimplication, but the following equivalences can be used:
$$
\begin{array}{r@{~~~}c@{~~~}l}
\top & \equiv & \bot \rightarrow \bot\\
\neg A & \equiv & A \rightarrow \bot\\
A \leftrightarrow B & \equiv & (A \rightarrow B) \land (B \rightarrow A)
\end{array}
$$
The symbols $A$ and $B$ are arbitrary formulas.  

\section{Natural Deduction in \nadea}

One of the unique features of \nadea\ is that it comes with a formalization in Isabelle of its proof system.

\subsection{Syntax for Terms and Formulas}

The terms and formulas of the first-order logic language are defined as the data types \x{term} and \x{formula} (later abbreviated \x{tm} and \x{fm}, respectively). The type \x{identifier} represents predicate and function symbols (later abbreviated \x{id}).
\begin{trivlist}
\item
\begin{tabular}{@{~~~~~~~~~~}l}
\x{identifier := string} \\[1ex]
\x{term := Var nat $\mid$ Fun identifier [term, ..., term]} \\[1ex]
\x{formula := Falsity $\mid$ Pre identifier [term, ..., term] $\mid$ Imp formula formula $\mid$} \\[.5ex] \phantom{\x{formula :=}} \x{Dis formula formula $\mid$ Con formula formula $\mid$ Exi formula $\mid$ Uni formula}
\end{tabular}
\end{trivlist}
Truth, negation and biimplication are abbreviations.
In the syntax of our formalization, we refer to variables by use of the de Bruijn indices. That is, instead of identifying a variable by use of a name, usually $x$, $y$, $z$ etc., each variable has an index that determines its scope. The use of de Bruijn indices instead of named variables allows for a simple definition of substitution. Furthermore, it also serves the purpose of teaching the students about de Bruijn indices.
Note that we are not advocating that de Bruijn indices replace the standard treatment of variables in general.
It arguably makes complex formulas harder to read, but the pedagogical advance is that the notion of scope is exercised.

\subsection{Natural Deduction Rules}

Provability in \nadea\ is defined inductively as follows:
\begin{trivlist}
\item
\begin{small}\label{dedRules}
$$
\infer[\x{Assume}]
{\x{OK p a}}{\x{member p a}}
~~~~~~~~~~~~~~~~
\infer[\x{Boole}]
{\x{OK p a}}{\x{OK Falsity ((Imp p Falsity) \# a )}}
$$

$$
\infer[\x{Imp\_E}]
{\x{OK q a}}{\x{OK (Imp p q) a} ~~&~~ \x{OK p a}}
~~~~~~~~~~~~~~~~
\infer[\x{Imp\_I}]
{\x{OK (Imp p q) a}}{\x{OK q (p \# a)}}
$$

$$
\infer[\x{Dis\_E}]
{\x{OK r a}}{\x{OK (Dis p q) a} ~~&~~ \x{OK r (p \# a)} ~~&~~ \x{OK r (q \# a)}}
$$

$$
\infer[\x{Dis\_I1}]
{\x{OK (Dis p q) a}}{\x{OK p a}}
~~~~~~~~~~~~~~~~
\infer[\x{Dis\_I2}]
{\x{OK (Dis p q) a}}{\x{OK q a}}
$$

$$
\infer[\x{Con\_E1}]
{\x{OK p a}}{\x{OK (Con p q) a}}
~~~~~~~~~~~~~~~~
\infer[\x{Con\_E2}]
{\x{OK q a}}{\x{OK (Con p q) a}}
~~~~~~~~~~~~~~~~
\infer[\x{Con\_I}]
{\x{OK (Con p q) a}}{\x{OK p a} ~~&~~ \x{OK q a}}
$$

$$
\infer[\x{Exi\_E}]
{\x{OK q a}}{\x{OK (Exi p) a} &~ \x{OK q ((sub 0 (Fun c []) p) \# a)} & \x{news c (p\#q\#a)}}
$$

$$
\infer[\x{Exi\_I}]
{\x{OK (Exi p) a}}{\x{OK (sub 0 t p) a}}
$$

$$
\infer[\x{Uni\_E}]
{\x{OK (sub 0 t p) a}}{\x{OK (Uni p)}}
~~~~~~~~~~~~~~~~
\infer[\x{Uni\_I}]
{\x{OK (Uni p) a}}{\x{OK (sub 0 (Fun c []) p) a} ~~&~~ \x{news c (p \# a))}}
$$
\end{small}
\end{trivlist}
\x{OK p a} means that the formula \x{p} follows from the list of assumptions \x{a} and \x{member p a} means that \x{p} is a member of \x{a}. The operator \x{\#} is between the head and the tail of a list. \x{news c l} checks if the identifier \x{c} does not occur in the any of the formulas in the list \x{l} and \x{sub n t p} returns the formula \x{p} where the term \x{t} has been substituted for the variable with the de Bruijn index \x{n}. Instead of writing \x{OK p a} we could also use the syntax \x{a} $\vdash$ \x{p}, even in Isabelle, but we prefer a more programming-like approach. In the types we use $\Rightarrow$ for function spaces. The definitions of \x{member}, \x{news} and \x{sub} are as follow:
\begin{trivlist}
\item
\begin{tabular}{@{~~~~~~~~~~}l}
\x{member :: fm $\Rightarrow$ fm list $\Rightarrow$ bool} \\
\x{member p [\,] = False} \\
\x{member p (q \# a) = (if p = q then True else member p a)} \\[0.5ex]
\x{new\_term :: id $\Rightarrow$ tm $\Rightarrow$ bool} \\
\x{new\_term c (Var v) = True} \\
\x{new\_term c (Fun i l) = (if i = c then False else new\_list c l)} \\[0.5ex]
\x{new\_list :: id $\Rightarrow$ tm list $\Rightarrow$ bool} \\
\x{new\_list c [\,] = True} \\
\x{new\_list c (t \# l) = (if new\_term c t then new\_list c l else False)} \\[0.5ex]
\x{new :: id $\Rightarrow$ fm $\Rightarrow$ bool} \\
\x{new c Falsity = True} \\
\x{new c (Pre i l) = new\_list c l} \\
\x{new c (\makebox[\conwidthx][l]{Imp} p q) = (if new c p then new c q else False)} \\
\x{new c (\makebox[\conwidthx][l]{Dis} p q) = (if new c p then new c q else False)} \\
\x{new c (\makebox[\conwidthx][l]{Con} p q) = (if new c p then new c q else False)} \\
\x{new c (\makebox[\uniwidthx][l]{Exi} p) = new c p} \\
\x{new c (\makebox[\uniwidthx][l]{Uni} p) = new c p}  \\[0.5ex]
\x{news :: id $\Rightarrow$ fm list $\Rightarrow$ bool} \\
\x{news c [\,] = True} \\
\x{news c (p \# a) = (if new c p then news c a else False)}
\end{tabular}
\end{trivlist}
\begin{trivlist}
\item
\begin{tabular}{@{~~~~~~~~~~}l}
\x{inc\_term :: tm $\Rightarrow$ tm} \\
\x{inc\_term (Var v) = Var (v + 1)} \\
\x{inc\_term (Fun i l) = Fun i (inc\_list l)} \\[0.5ex]
\x{inc\_list :: tm list $\Rightarrow$ tm list} \\
\x{inc\_list [\,] = [\,]} \\
\x{inc\_list (t \# l) = inc\_term t \# inc\_list l} \\[0.5ex]
\x{sub\_term :: nat $\Rightarrow$ tm $\Rightarrow$ tm $\Rightarrow$ tm} \\
\x{sub\_term n s (Var v) = (if v = n then s else if v $>$ n then Var (v -- 1) else Var v)} \\
\x{sub\_term n s (Fun i l) = Fun i (sub\_list n s l)} \\[0.5ex]
\x{sub\_list :: nat $\Rightarrow$ tm $\Rightarrow$ tm list $\Rightarrow$ tm list} \\
\x{sub\_list n s [\,] = [\,]} \\
\x{sub\_list n s (t \# l) = sub\_term n s t \# sub\_list n s l} \\[0.5ex]
\x{sub :: nat $\Rightarrow$ tm $\Rightarrow$ fm $\Rightarrow$ fm} \\
\x{sub n s Falsity = Falsity} \\
\x{sub n s (Pre i l) = Pre i (sub\_list n s l)} \\
\x{sub n s (\makebox[\conwidthx][l]{Imp} p q) = \makebox[\conwidthx][l]{Imp} (sub n s p) (sub n s q)} \\
\x{sub n s (\makebox[\conwidthx][l]{Dis} p q) = \makebox[\conwidthx][l]{Dis} (sub n s p) (sub n s q)} \\
\x{sub n s (\makebox[\conwidthx][l]{Con} p q) = \makebox[\conwidthx][l]{Con} (sub n s p) (sub n s q)} \\
\x{sub n s (\makebox[\uniwidthx][l]{Exi} p) = \makebox[\uniwidthx][l]{Exi} (sub (n + 1) (inc\_term s) p)} \\
\x{sub n s (\makebox[\uniwidthx][l]{Uni} p) = \makebox[\uniwidthx][l]{Uni} (sub (n + 1) (inc\_term s) p)}
\end{tabular}
\end{trivlist}

\subsection{Semantics for Terms and Formulas}

To give meaning to formulas and to prove \nadea\ sound we need a semantics of the first-order logic language. This semantics is defined in the formalization in Isabelle, and it is thus not part of the tool itself. We present the semantics below. \x{e} is the environment, i.e. a mapping of variables to elements. \x{f} maps function symbols to the maps they represent. These maps are from lists of elements of the universe to elements of the universe. Likewise, \x{g} maps predicate symbols to the maps they represent. \x{'u} is a type variable that represents the universe. In can be instantiated with any type. For instance, it can be instantiated with the natural numbers, the real number or strings.
\begin{trivlist}
\item
\begin{tabular}{@{~~~~~~~~~~}l}
\x{semantics\_term :: (nat $\Rightarrow$ 'u) $\Rightarrow$ (id $\Rightarrow$ 'u list $\Rightarrow$ 'u) $\Rightarrow$ tm $\Rightarrow$ 'u} \\
\x{semantics\_term e f (Var v) = e v} \\
\x{semantics\_term e f (Fun i l) = f i (semantics\_list e f l)} \\[1ex]
\x{semantics\_list :: (nat $\Rightarrow$ 'u) $\Rightarrow$ (id $\Rightarrow$ 'u list $\Rightarrow$ 'u) $\Rightarrow$ tm list $\Rightarrow$ 'u list} \\
\x{semantics\_list e f [\,] = [\,]} \\
\x{semantics\_list e f (t \# l) = semantics\_term e f t \# semantics\_list e f l} \\[1ex]
\x{semantics ::  (nat $\Rightarrow$ 'u) $\Rightarrow$ (id $\Rightarrow$ 'u list $\Rightarrow$ 'u) $\Rightarrow$ (id $\Rightarrow$ 'u list $\Rightarrow$ bool) $\Rightarrow$ fm $\Rightarrow$ bool} \\
\x{semantics e f g Falsity = False} \\
\x{semantics e f g (Pre i l) = g i (semantics\_list e f l)} \\
\x{semantics e f g (\makebox[\conwidthx][l]{Imp} p q) = (if semantics e f g p then semantics e f g q else True)} \\
\x{semantics e f g (\makebox[\conwidthx][l]{Dis} p q) = (if semantics e f g p then True else semantics e f g q)} \\
\x{semantics e f g (\makebox[\conwidthx][l]{Con} p q) = (if semantics e f g p then semantics e f g q else False)} \\
\x{semantics e f g (\makebox[\uniwidthx][l]{Exi} p) = (\makebox[\qmwidthx][l]{?} x. semantics (\% n. if n = 0 then x else e (n -- 1)) f g p)} \\
\x{semantics e f g (\makebox[\uniwidthx][l]{Uni} p) = (\makebox[\qmwidthx][l]{!} x. semantics (\% n. if n = 0 then x else e (n -- 1)) f g p)}
\end{tabular}
\end{trivlist}
Most of the cases of \x{semantics} should be self-explanatory, but the \x{Uni} case is complicated. The details are not important here, but in the case for \x{Uni} it uses the universal quantifier (\x{!}) of Isabelle's higher-order logic to consider all values of the universe. It also uses the lambda abstraction operator (\x{\%}) to keep track of the indices of the variables. Likewise, the case for \x{Exi} uses the existential quantifier (\x{?}) of Isabelle's higher-order logic.

We have proved soundness of the formalization in Isabelle (shown here as a derived rule):
$$
\infer[\text{Soundness}]{\x{semantics e f g p}}{\x{OK p [\,]}}
$$
This result makes \nadea\ interesting to a broader audience since it gives confidence in the formulas proved using the tool.

\section{Construction of a Proof}

We now describe the core features of \nadea\ from the perspective of the user. That is, we uncover how to use \nadea\ to conduct and edit a proof as well as how proofs are presented.

In order to start a proof, you have to start by specifying the goal formula, that is, the formula you wish to prove. To do so, you must enable editing mode by clicking the Edit button in the top menu bar. This will show the underlying proof code and you can build formulas by clicking the red \textcurrency~symbol. Alternatively, you can load a number of tests by clicking the Load button.

At all times, once you have fully specified the conclusion of any given rule, you can continue the proof by selecting the next rule to apply. Again you can do this by clicking the the red \textcurrency~symbol. Furthermore, \nadea\ allows for undoing and redoing editing steps with no limits.

All proofs are conducted in backward-chaining mode. That is, you must start by specifying the formula that you wish to prove. You then apply the rules inductively until you reach a proof --- if you can find one. The proof is finished by automatic application of the \x{Assume} rule once the conclusion of a rule is found in the list of assumptions.

To start over on a new proof, you can load the blank proof by using the Load button, or you can refresh the page. Please note that any unsaved work will then be gone.

In \nadea\ we present any given natural deduction proof (or an attempt at one) in two different types of syntax. One syntax follows the rules as defined in section \ref{dedRules} and is closely related to the formalization in Isabelle, but with a redefined and more simple syntax in terms of learning. The proof is not built as most often seen in the literature about natural deduction. Usually, for each rule the premises are placed above its conclusion separated by a line. We instead follow the procedure of placing each premise of the rule on separate lines below its conclusion with an additional level of indentation.

\noindent
\begin{minipage}[t]{.62\textwidth}
\vspace{0pt}
\includegraphics[scale=.5]{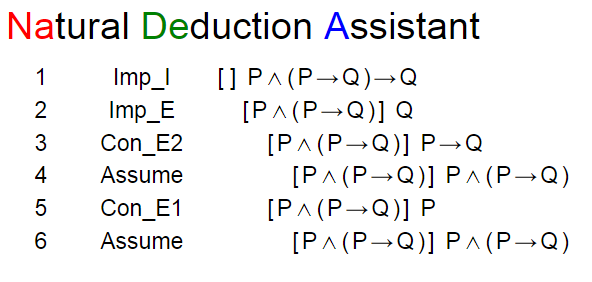}
\end{minipage}
\begin{minipage}[t]{.37\textwidth}
\vspace{0pt}
\vspace{4ex}
$$
\infer[^{(1)}]
{p \land (p \rightarrow q) \rightarrow q}
{
	\infer
    {q}
    {
    	\infer
        {p \rightarrow q}
        {
        	\infer[^{(1)}]
            {p \land (p \rightarrow q)}
            {}
        }
    	&
    	\infer
        {p}
        {
        	\infer[^{(1)}]
            {p \land (p \rightarrow q)}
            {}
        }
    }
}
$$
\end{minipage}
The above proof also can be written in terms of the \x{OK} syntax as follows:
$$
\begin{array}{@{}lll@{}}
\x{1} & \x{OK (Imp (Con (Pre "P" []) (Imp (Pre "P" []) (Pre "Q" []))) (Pre "Q" [])) []} & \x{Imp\_I}
\\[1ex]
\x{2} & \x{~~ OK (Pre "Q" []) [(Con (Pre "P" []) (Imp (Pre "P" []) (Pre "Q" [])))]} & \x{Imp\_E}
\\[1ex]
\x{3} & \x{~~~~ OK (Imp (Pre "P" []) (Pre "Q" []))}
\\ & \x{~~~~~~~~~~\, [(Con (Pre "P" []) (Imp (Pre "P" []) (Pre "Q" [])))]} & \x{Con\_E2}
\\[1ex]
\x{4} & \x{~~~~~~ OK (Con (Pre "P" []) (Imp (Pre "P" []) (Pre "Q" [])))}
\\ & \x{~~~~~~~~~~\, [(Con (Pre "P" []) (Imp (Pre "P" []) (Pre "Q" [])))]} & \x{Assume}
\\[1ex]
\x{5} & \x{~~~~ OK (Pre "P" []) [Con (Pre "P" []) (Imp (Pre "P" []) (Pre "Q" [])))]} & \x{Con\_E1}
\\[1ex]
\x{6} & \x{~~~~~~ OK (Con (Pre "P" []) (Imp (Pre "P" []) (Pre "Q" [])))}
\\ & \x{~~~~~~~~~~\, [(Con (Pre "P" []) (Imp (Pre "P" []) (Pre "Q" [])))]} & \x{Assume}
\end{array}
$$

\section{Related Work}

Throughout the development of \nadea\ we have considered some of the natural deduction assistants currently available. Several of the tools available share some common flaws. They can be hard to get started with, or depend on a specific platform.
However, there are also many tools that each bring something useful and unique to the table. One of the most prominent is \textsc{Panda}, described in \cite{panda}. \textsc{Panda} includes a lot of graphical features that make it fast for the experienced user to conduct proofs, and it helps the beginners to tread safely. Another characteristic of \textsc{Panda} is the possibility to edit proofs partially before combining them into a whole. It definitely serves well to reduce the confusion and complexity involved in conducting large proofs. However, we still believe that the way of presenting the proof can be more explicit. In \nadea, every detail is clearly stated as part of the proof code. In that sense, the students should become more aware of the side conditions to rules and how they work.

Another tool that deserves mention is ProofWeb \cite{proofweb} which is open source software for teaching natural deduction. It provides interaction between some proof assistants (Coq, Isabelle, Lego) and a web interface.
The tool is highly advanced in its features and uses its own syntax. Also, it gives the user the possibility to display the proof in different formats. However, the advanced features come at the cost of being very complex for undergraduate students and require that you learn a new syntax. It serves as a great tool for anyone familiar with natural deduction that wants to conduct complex proofs that can be verified by the system. It may, on the other hand, prove less useful for teaching natural deduction to beginners since there is no easy way to get started. In \nadea, you are free to apply any (applicable) rule to a given formula, and thus, beginners have the freedom to play around with the proof system in a safe way. Furthermore, the formalized soundness result for the proof system of \nadea\ makes it relevant for a broader audience, since this gives confidence in that the formulas proved with the tool are actually valid.

\section{Further Work}

In \nadea\ there is support for proofs in propositional logic as well as first-order logic. We would also like to extend to more complex logic languages, the most natural step being higher-order logic. This could be achieved using the CakeML approach \cite{Kumar}.
Other branches of logic would also be interesting, and the possibilities are numerous. Apart from just extending the natural deduction proof system to support other types of logic, another option is to implement other proof systems as well.

Because the \nadea\ tool has a formalization in Isabelle of its proof system, we would like to provide features that allow for a direct integration with Isabelle. For instance, we would like to allow for proofs to be exported to an Isabelle format that could verify the correctness of the proofs.
A formal verification of the implementation would require much effort, but perhaps it could be reimplemented on top of Isabelle (although probably not in TypeScript / JavaScript).

We would like to extend \nadea\ with more features in order to help the user in conducting proofs and in understanding logic. For example, the tool could be extended with step-by-step execution of the auxiliary primitive recursive functions used in the side conditions of the natural deduction rules.

So far only a small group of computer science students have tested \nadea, but it will be classroom tested with around 60 bachelor students in the next semester.
Currently the tool has no support for student assignments and automatic feedback and/or grading.
The tool could be extended such that the students are evaluated and perhaps given a score based on the proofs they conduct. It is not obvious how this could best be implemented.
We hope to find the resources for the development of such features but already now we think that the tool has the potential to be one of the main ways to teach logic in mathematics and computer science.

\subparagraph*{Acknowledgements}

We would like to thank Stefan Berghofer for discussions about the formalization of natural deduction in Isabelle.
We would also like to thank Andreas Halkj{\ae}r From and Andreas Viktor Hess for comments on the paper.

\end{document}